\documentclass[pdflatex, sn-chicago]{sn-jnl}

\usepackage{graphicx}%
\usepackage{multirow}%
\usepackage{amsmath,amssymb,amsfonts}%
\usepackage{amsthm}%
\usepackage{mathrsfs}%
\usepackage[title]{appendix}%
\usepackage{xcolor}%
\usepackage{textcomp}%
\usepackage{manyfoot}%
\usepackage{tabto}
\usepackage{algorithm}%
\usepackage{algorithmicx}%
\usepackage{algpseudocode}%
\usepackage{listings}%

\usepackage[textsize=tiny]{todonotes}

\theoremstyle{thmstyleone}%
\newcommand{\GH}{\vspace{\baselineskip}\par G. H.  \tabto{20mm}}

\newcommand{\YA}{\vspace{\baselineskip}\par Y. A. \tabto{20mm}}

\newcommand{\schrodinger}{Schr\"{o}dinger}

\begin{document}

\title[A Personal History in Quantum Foundations]{A Personal History in Quantum Foundations: \newline an Oral History Interview$^{\ddagger}$}

\author[1,2]{\fnm{Yakir} \sur{Aharonov}}

\author*[3,4]{\fnm{Guy} \sur{Hetzroni}}\email{guyhetzroni@humanities.haifa.ac.il}

\affil[1]{\small\orgdiv{Raymond and Beverly Sackler School of Physics
and Astronomy}, \linebreak \orgname{Tel Aviv University}, \orgaddress{\city{Tel Aviv}, \postcode{69978}, \country{Israel}}}

\affil[2]{\small\orgdiv{Institute for Quantum Studies}, \orgname{Chapman University}, \orgaddress{\city{Orange, CA}, \postcode{92866}, \country{USA}}}

\affil[3]{\small\orgdiv{Department of Natural Sciences}, \orgname{The Open University of Israel}, \linebreak \orgaddress{\city{Ra'anana}, \postcode{4353701}, \country{Israel}}}

\affil[4]{\small\orgdiv{Department of Philosophy}, \orgname{University of Haifa}, \orgaddress{\city{Haifa}, \postcode{3103301}, \country{Israel}}}

\abstract{This oral history interview provides Yakir Aharonov's personal and conceptual recollections from the 1950s to the 2000s, focusing on his academic trajectory, some of his major contributions to quantum foundations, and his perspective on the development of the field.  The interview traces Aharonov's academic career through the Technion, University of Bristol, Brandeis University, Yeshiva University, University of South Carolina, Tel Aviv University, George Mason University, and Chapman University. The interview also traces the development of Aharonov's work and collaborations on the EPR paradox, time-energy uncertainty relations, quantum measurement theory, quantum frames of reference, the two-state vector formalism, weak values, protective measurements, and several other contributions. It includes recollections of meetings and discussions with David Bohm, Niels Bohr, John Clauser, Paul Dirac, Rudolf Peierls, Aage Petersen, and Abner Shimony. Also mentioned: David Albert, Jeeva Anandan, Peter Bergmann, Gideon Carmi, Avshalom Elitzur, Dave Finkelstein, Vladimir Fock, Itzhak Kalugai, Ed Lerner, Joel Lebowitz, Stanley Mandelstam, Yuval Ne’eman, Sandu Popescu, Maurice Pryce, Daniel Rohrlich, Nathan Rosen, Joseph Safko, Paul Singer, Daniele Struppa, Leonard Susskind, Lev Vaidman, Chien-Shiung Wu.
}

\keywords{Yakir Aharonov, David Bohm, Niels Bohr, quantum foundations, EPR paradox, weak measurements, quantum frames of reference.}

{\let\thefootnote\relax\footnotetext{$^{\ddagger}$Preprint of a paper forthcoming in \emph{The European Physical Journal H}; please cite the published version.}}
\maketitle

\section{Personal background and studies at the Technion}

\GH  Can you describe your upbringing and family background?\footnote{The interview  was conducted in a series of meetings in Tel Aviv during 2025--2026. Edits of language, structure, and clarity were made on Aharonov's request. All footnotes were added by G. H. Note that the discovery of the Aharonov-Bohm effect, the debates about it, and related issues (e.g., modular momentum) were covered separately in \citet{aharonov2025hetzroni}. }

\YA Both my parents immigrated to Israel in the early 1920s: my mother came from Lithuania and my father from Russia. Both were Zionists. They met and got married here, in Israel. I was born in Haifa in 1932 and when I was three years old, my family moved to Kiryat Haim, which was just established, all houses had land around them.  We grew all kinds of trees and vegetables and had chickens. It was a very nice atmosphere. So I grew up in Kiryat Haim and went there to elementary school and high school. 

As a child I liked playing the violin, I learned for about 10 years. As a joke, when I practised I found that the best acoustics in the house were in the kitchen or the bathroom. I later read that Einstein discovered the same thing. I continued to play chamber music with some friends as a hobby maybe until I was 50.  

When I was 17 years old, I became very interested in chess, and for one year, I played a lot of chess, and at that time I was the most promising young chess player in Israel, but I stopped playing professionally after I found it too absorbing, it really occupied all my mind.\footnote{On Aharonov's early biography see \citet{pines2014yakir}. } 

\GH What do you remember from your high school? 

\YA In high school I was quite a mediocre student. I hated physics because of the way it was taught, just by memorizing formulas. The only thing I was really good at was mathematics. I remember my parents came to meet my math teacher. His name was Kurz, and he told my parents, `you should make sure that your son does not study mathematics, because he will end up like myself. Better make sure that he will study engineering or something like that.'

I read a lot of philosophy at the time. Western philosophy, Eastern philosophy... I thought that philosophers knew how to ask good questions, but they don't know how to answer them. I became more interested in physics after I encountered \citet{eddington1928natureHEB} and \citet{einstein1939infeldHEB}. I saw that physics helped Eddington address philosophical questions in a better way than the philosophers who had previously explored these areas. So I decided to study theoretical physics. Before that, I had to spend two years in the army, 1950 to 1952, and then I went to the Technion. 

I studied for my bachelor's degree at the Technion from '52 to '56, in the early days of its Faculty of Sciences.

\GH Can you describe the physics community at the Technion? 

\YA There was a science faculty. There were some physicists, I can hardly remember the names. Nathan Rosen was the main force behind the idea of a science faculty, before him there was mainly engineering. There were also some chemists. Apart from Rosen, there was little research going on. People in the faculty immigrated from different places. They had PhDs in physics, but were mainly engaged in teaching. I do remember Franz Ollendorff, he was known in engineering for his contributions. 

I had a teacher, a professor in chemistry, by the name of Itzhak Kalugai. On the first day of his class, he said, `you know, there are some people claiming that the way to understand chemistry is through this theory they call quantum mechanics. I don't believe in it. I will teach you chemistry as it should be taught'. All his classes were just memorizing things. I would have failed the course if I didn't go the restroom to copy the answers.

\GH Do you remember any particular classmates?

\YA I remember that Paul Singer studied physics with me. And Yona Oren. Both of them later became professors, and Paul Singer had a very important position in the Israel Science Foundation. Altogether we were about seven students in physics. Also, Asher Peres was studying at the same time, but he was studying engineering, he switched to physics only later on.   

\GH What else do you remember from your time at the Technion? 

\YA I had to work during the time I studied there because my parents could not really support me. I taught English and also some algebra in evening school. During the summer breaks I did officer's training in the IDF.  

\section{Working with David Bohm}

\GH In our previous interview \citep{aharonov2025hetzroni} you told me how you started working with Bohm in 1956, while he was at the Technion and you were towards the end of your undergraduate studies, and how this led to you going with Bohm to do your PhD at the University of Bristol and to the Aharonov-Bohm effect. But you also wrote other significant papers together. 

\YA The first article we published \citep{aharonov1957velocity} was actually not very important. It was connected with some comment that Dirac made, that in his equation the velocity of the electron must be either $+c$ or $-c$. The first thing I told Bohm was that I don't like this comment, and we argued against it. After we decided that I would join Bohm in moving to Bristol to do my PhD we started to work more seriously together. 

\GH The Sinai war broke at the end of October 1956 when you were still in Israel. What do you recall from it?

\YA Yes. I served as an artillery officer and was sent to the northern border. I was helping to develop a system that would allow for spotting the enemy's guns. The idea was to put a number of microphones and then to locate the artillery of the enemy based on the time differences in the reception time.

Anyway, eventually Syria did not enter the war. So I was there for a small number of weeks, and I had with me the book by \citet{vonNeumann1955principles}, `Mathematical Foundations of Quantum Mechanics'. I studied it very carefully during these weeks. It influenced my later work. 

\GH And then, soon after, you were a PhD student in Bristol. 
 
\YA I spent three years in Bristol and it was a very fruitful time. My dissertation \citep{aharonov1960phd} was about the quantum theory of measurements. In one part we discussed the EPR paper. Another part was based on an idea I had on time-energy uncertainty relations. It is about the possibility to perform measurements of the energy in arbitrarily short times and the significance of relativity for quantum measurement. The other part was the Aharonov-Bohm effect. 

\GH Your paper on EPR \citep{aharonov1957epr} immediately turned the suggestion of \citeauthor{epr1935} (\citeyear{epr1935}) from a thought experiment not merely into a feasible experiment, but in fact into one that had already been performed seven years earlier by \citet{wu1950annihilation}.\footnote{On Wu's work in the context of quantum foundations see \citet{silva2022wu}. } The experiment was actually suggested by \citet{wheeler1946polyelectrons} as a test for pair theory. How did Bohm and yourself come to work on that, and how did your collaboration go? 

\YA That was Bohm's idea. The only contribution I made to this article was the calculation, which was important, it wasn't easy, but the ideas were Bohm's through that phase. 

In her experiment, Wu checked the disintegration of positronium into two photons that go in different directions, and she checked the correlation of their polarizations, to find out whether their correlation matches the original zero relative angular momentum state of the positronium. Bohm suggested to check the correlation to see whether it could be explained classically in terms of two photons that come with definite opposite polarizations in random different directions. We checked, and when I did the calculation we saw that you can't explain this classically. So this was much before Bell. It was actually the first, real experimental test to show that quantum correlations go beyond classical ones. Bell later proved that it's general, but this was the first particular example of quantum correlation that are stronger than what you can achieve classically. 

By the way, this article and the later comment about it \citep{aharonov1960epr2} are the only articles we wrote where Bohm's name appears first, because it was his idea. The other articles we wrote together were my idea.

\GH John \citet{bell1964} indeed cites your paper. Can you elaborate in what way your paper anticipated his results? 

\YA We did not show that any classical explanation of the correlations is impossible, but we showed that the most natural classical explanation fails.

\GH Your paper mentions private communication with Einstein, in which he suggested the the `current formulation of the many-body problem in quantum mechanics may break down when particles are far enough apart' (p. 1071). Do you remember Bohm discussing his communication with Einstein?

\YA No.

\GH Interestingly, your paper also notes that this suggestion was already developed in responses to the EPR by Wendell \citet{furry1936eprA,furry1936eprB}, who suggested a reduction to a product state as an alternative to the entangled state. Did this citation also come from Bohm? 

\YA Yes. There might have been some relation between Bohm and Furry,  because both of them were harmed by McCarthyism. So maybe there was some contact between them, but I'm not sure. I met Furry later\footnote{See in \citet{aharonov2025hetzroni}, \S3. }, by the way, but I don't remember talking to him about Bohm.

\GH Did anybody think of connecting this experiment to quantum foundations before you, or to what today we call entanglement? 

\YA No. Bohm was the first one who thought about this experiment as a way to check the EPR, not entanglement, but the EPR idea. 

\GH Do you have an explanation for why no one saw this connection before? 

\YA I think that before our paper people did not think that correlations between polarizations had anything to do with EPR. Before Bohm came up with the idea, people only thought about the EPR in the original language of position and momentum. Nobody thought to generalize it to other variables before Bohm. Earlier \citep{bohm1951quantum} he suggested doing the experiment with spin, and in our paper with photons. 

\GH One widely held view is that the reason physicists did not recognize the significance of entanglement at that stage is related to the dominance of the Copenhagen interpretation. For someone who experienced it first-hand, does it make sense to you that this is also the case here?

\YA I think at that time people were so confused about quantum mechanics that they were just able to follow the formalism. And when it came to discussion of interpretation, it was always the Copenhagen idea of complementarity. And I don't think that people really thought much about how to connect complementarity with variables like spin or polarization. In fact, it was well known at the time that Bohr even argued that you can't measure the spin of an electron.\footnote{On that, see \citet{morrison2007spin}. } The Stern-Gerlach experiment was conducted with neutral atoms, because with electrons the electromagnetic interaction with the charge completely masks the interaction with the spin. So people didn't really think  carefully about an electron by itself, or a photon by itself, having these properties. 

\GH Did Bohm object to this view of spin by Bohr? 

\YA Bohm did not directly refer to this at all. He certainly talks about electrons as having spin, and of photons as having polarization. He was very clear about what quantum mechanics says about these properties, he wasn't confused.

\GH Do you know how Bohm heard on the Wu-Shaknov experiment at that particular time? It is known that he was in touch in 1957 with Maurice Pryce, who had previously written about it \citep{pryce1947annihilation}, and on the other hand, Wu had just published her famous paper on violation of parity conservation  \citep{wu1957parity} which must have made waves. 

\YA I don't know. He just came to me and said `look, here is a way that we can check whether quantum mechanics is right or wrong in this situation.'

\GH During World War II both David Bohm and Chien-Shiung Wu were at Berkeley Radiation Laboratory. Do you know if Bohm knew her? And did you or he ever discuss the experiment with Wu?  

\YA No. He never mentioned that to me. I don't think that he had a discussion with her. I never heard about it if he did. I never met her, also not later. 

\GH Your paper seems also to have been motivated in part by Bohm's theory \citep{bohm1952interpretation}. In particular, your paper says: `there has been developed a further new explanation of the quantum theory in terms of a deeper subquantum-mechanical level. The laws of this lower level are different from those of the quantum theory, but approach these latter laws as an approximation [...] Explanations of this kind will be published later. It will be seen with the aid of this theory that the paradox can be understood in a perfectly rational way, in terms of a new notion of coordinated fluctuations arising in the subquantum-mechanical level.' \citep[p. 1072]{aharonov1957epr}. But it was not published later.

\YA I think at that time Bohm had some idea how his hidden variable theory could develop to something different from ordinary quantum mechanics. Bohm was hoping that there is a hidden variable theory with only local interactions, but this was not something that we were able to talk about with each other. I remember that Bohm was hoping that perhaps something would be wrong with the quantum predictions. But it was, of course, wrong, our results showed that quantum mechanics was right.

\GH Then, there was your work on the time-energy uncertainty relations that was part of your thesis and published in \cite{aharonov1961uncertainty}. 

\YA There are two things there. One is the time–energy uncertainty, and the other is the claim that you can’t measure momentum very quickly due to special relativity. 

The time–energy uncertainty was the beginning. I told Bohm that the time at which you measure something has to do with the clock in the laboratory, it has nothing to do with the system itself. I said that the only meaningful time–energy uncertainty for a system is related to how the system itself defines time through its own evolution.

\GH Your paper targets a view expressed by \citet{landau1931messungsgrossen} and \citet{fock1947uncertaintyrelation}.
\YA  It was a wrong view held by many people. Niels Bohr himself wrote an article, in the context of collisions, saying that if you try to measure the energy at a short time then 
you necessarily introduce a large uncertainty in that energy \citep{bohr1928quantumpostulate}. 

Now, you can certainly measure the energy of a system using an external time parameter. The basic idea was to use von Neumann’s measurement scheme. Let's say the position $Q$ of the pointer is coupled to the momentum $p$ of the particle that is being measured. The interaction Hamiltonian is $H_m=g(t)Qp$, with $g(t)$ vanishing except for during an arbitrarily short time interval, at the cost of having the maximal value of $g(t)$ sufficiently large. At the end, the momentum of the pointer is correlated with the momentum of the particle, so you can measure the kinetic energy in this way.  The velocity  of the particle is $\frac{\partial H}{\partial p}=p/m+g(t)Q$. It can become highly uncertain for the arbitrarily short duration of the measurement, but then it returns to its original value.  

Then there was another issue, that of connecting measurement with the finite velocity of light, and showing that you can’t measure energy or momentum in a very short time if you take the speed of light into account. To measure momentum, you must introduce a large uncertainty in position, and if velocity is limited, it takes time to create that uncertainty. That was the issue I tried to discuss with Bohr (\S\ref{bohr}). 

\GH There was much objection to the claim that you can measure the energy in arbitrarily short time, including an exchange with Fock \citep{fock1962timeenergy,aharonov1964timeenergy,fock1966aharonov}. Fock presented his view as a defence of quantum mechanics against an attempt to reject the Heisenberg-Bohr uncertainty relations.  That exchange did not seem to converge to an agreement. You also later gave a more general argument for your point in \citet{aharonov1975noncanonical}. What do you remember from this debate? Did you ever meet Vladimir Fock, or discuss the issue with him beyond the papers? 

\YA No, I never met Fock. I must say that I have extremely limited recollection of the details of this debate. He was persistent, but he was completely wrong.

But I remember other objections. Rudolf Peierls was at the time the chairman of the physics department at Birmingham University. After Dirac he was the most famous physicist in England. At the end of my PhD, he was my external examiner. During the examination Bohm wasn’t allowed to be in the room. It was only me with Maurice Pryce and Peierls, my two examiners. So I remember I entered   this room frightened, and Peierls, a small guy, said `well... I have to tell you something. The two of us, Pryce and I, are certain that what you wrote about energy and time is wrong. But we sat all morning'--- it was Saturday afternoon---`to find the mistake in what you wrote, and we could not find it. We are still sure it’s wrong but we couldn’t find the mistake. And anyhow we like very much what you wrote in the other parts of the thesis. You’ll get your doctorate, but remember we don’t believe it.' 

A few weeks later I got a call from Birmingham. It was Stanley Mandelstam, who was Peierls’s assistant. He told me that Peierls wants me to come to discuss it some more. So I came and I stayed there for a few days. He still thought that you can’t measure the energy in a short time, and I showed him an example where you could do that, something from my thesis, and we discussed it for quite a while. Finally he said that he agreed with me. 

By the way, many years later, maybe twenty, there was a meeting in Israel, a memorial conference for Lev Landau.\footnote{That was most probably the 1988 Landau Memorial Conference on Frontiers of Physics, see \citet{rechenberg1988landaumemorial}. } Peierls was invited. So I was walking there in the corridor and Peierls came walking in the opposite direction and said: `Aharonov?! I don’t believe you again!' I think he died without realizing he was wrong. Now, of course, everybody accepts my view.  

\GH In 1960 you finished your PhD and moved to the United States, and later to Israel. Did you continue to be in touch with David Bohm? 

\YA Very, very little, because at that time Bohm was becoming more and more involved in developing his hidden variable theory and his work became increasingly philosophical. We met maybe once every few years. I remember the last time we met was in Israel, when we both received an honorary doctorate from the Technion, and then I told him about my two-vector theory and he got excited about it. But in general we had very little interaction.

\section{On Niels Bohr}
\label{bohr}

\GH It seems that both the EPR paper in your PhD and your work on the energy-time uncertainty relations were in tension with the Bohrian view of quantum mechanics. You also told me that Bohr initially objected to the Aharonov-Bohm effect \citep[][\S3]{aharonov2025hetzroni}. Interestingly, you got to meet Bohr. 

\YA Yes. When I was still a PhD student Bohm was invited to the Institute for Theoretical Physics in Copenhagen and he took me along.
When I was there, I met Aage Petersen in the corridor, he was the assistant of Bohr, who would come much later to work with me in the United States. He asked me what I was working on, and I told him about the idea that you can't measure momentum over a short period of time. He said, `You know, I think Bohr will be interested in it, I will arrange a meeting between you and Bohr'. 

So indeed, a meeting was arranged. It was a small room. Bohm was there, and me, and Bohr... Bohm was sitting on the side, and Bohr and myself were sitting near the blackboard.
I said `Professor Bohr, I would like to tell you about a problem I have with measurement theory'. Bohr stopped me, and said `Let me tell you something'. He then gave me a spiel, for one hour, with all his ideas about measurement theory. And then he said `that was a very interesting conversation!'. 
And that was it. 

I was amazed. I remember I walked out startled and people asked me `What happened?' I told them what happened. And they said, `Well, at least you're in good company'. It turned out that a few years earlier Richard Feynman came to tell Bohr about his ideas of path integrals. He started saying `I want to superpose a particle going one way or another way'. Bohr stopped him and gave him a one-hour lecture on why you should not discuss orbits in quantum mechanics. And that was the end of that meeting. So I felt I was in good company.

\GH Do you remember what Bohr was talking about during that one-hour spiel?

\YA His usual ideas about measurement theory... how he solved some paradoxes. Nothing I didn’t already know. 

\GH Later in the 1960s, when Petersen was working with you, did you talk about Niels Bohr? 

\YA We had many discussions about Niels Bohr. Petersen had a lot of criticism of Bohr. He said that Bohr spent enormous effort rewriting things again and again, trying to find the right wording. Petersen eventually felt unhappy working with him because he did not see real progress over all those years.

\GH What was your view of Bohr's approach to quantum mechanics and its influence? 

\YA I think that many common things  people say about quantum mechanics are wrong: that indeterminism is a fundamental axiom. That there is no quantum reality, because every measurement creates disturbance, and that there is no logic. These wrong interpretations started with the way that Bohr was talking about quantum mechanics. All the thought experiments that were done at the beginning of quantum mechanics were  only designed to check its consistency, and not to find the real experiments that will show what is new in quantum mechanics.
An experiment aimed to answer this question should be an experiment that classical physicists would never think about. To put a solenoid between two slits, where the particle doesn’t go through. Or to impose on an experiment two boundary conditions in time: in its past and in its future. These are  experiments that I suggested that are meant to really show what is new in quantum mechanics.

Bohr, more or less, said that there is no way to understand quantum mechanics. Don't attempt to understand it, accept it. One example of how this approach was wrong and misleading, was the work of Bohr and Rosenfeld, who spent two or three years to find a way to measure the electric field in quantum mechanics. Later, at the University of South Carolina in \citet{aharonov1975noncanonical} we found that if you just follow carefully von Neumann's scheme of measurement, you get the right  result immediately and clearly, the one that took them two years to find. So it exactly shows how the refusal of Bohr to think about the mathematics of quantum theory as a story hindered him from finding a solution you can get in five minutes. It shows how unproductive  his approach to quantum mechanics was. 

In hindsight I also think Einstein was right and Bohr was wrong on the issue of uncertainty. Of course, neither of them realized that you can perform measurements that do not disturb the system. That insight eliminates the whole idea of complementarity and disturbance as fundamental principles. In my view these ideas gave rise to a lot of confusion. 

\section{Brandeis, Yeshiva, South Carolina, and Tel Aviv}
\label{sixties-seventies}

\GH How did you move from Bristol to Brandeis University?

\YA After my time at Bristol, in 1960, Bohm was invited to Brandeis for a year and he arranged for me and Gideon Carmi to join him. But eventually he wasn’t allowed into the States because he lost his American passport during the time he was in Brazil, where he went following the McCarthy era.\footnote{For more details on Bohm's passport and American citizenship see  \citet{freire2019bohm}, \S4.1. } Eventually he was allowed to visit the States, but only much later when his father died.

I spent a year at Brandeis, and then I came to Yeshiva in nineteen sixty-one. 

\GH Can you describe the move to Yeshiva University? 

\YA I remember that Dave Finkelstein came to meet me when I was still a research associate at Brandeis University. At that time I was wondering whether to stay at Brandeis, which I was offered to do. But then Finkelstein came there and convinced me to come to Yeshiva, telling me that the graduate school was still at an early stage, and that I could play a role in developing it. He mentioned some of the names of people that were there apart from him, like Joel Lebowitz and Peter Bergmann... Who else was there...?

\GH In `The Cosmic Landscape', \citet{susskind2006cosmiclandscape} vividly describes his arrival in Yeshiva University, mentioning walking into a big room with `the most brilliant collection of physicists that I had ever seen assembled in one place' (Chp. 2), explicitly mentioning Dave Finkelstein, Paul Dirac, Roger Penrose, Joel Lebowitz, Elliot Lieb and yourself. 

\YA Penrose? Maybe, I don’t remember. But anyway, I decided to accept the offer and came there, and it was a very, very good atmosphere. I was free to think about whatever I wanted and I had people to talk to. 

\GH Dirac was there during 1963--1964. Did you interact with him?  

\YA I attended his talks... He gave a series of lectures there, in which he was claiming that in quantum field theory you can’t have the \schrodinger\ representation, you can have only the Heisenberg representation. 

I also remember one  conversation with him.  I told him that if there is an array of solenoids, and you send a particle with a momentum that is low enough it will be reflected back. Suppose the solenoids each have half a fluxon, then the minimum momentum they can give the particle is $h/2L$, where L is the distance between the solenoids. If that momentum is bigger than the original momentum of the particle, the particle can’t get it. So the particle must be reflected back.
He told me that it looks funny that the particle will be reflected. On the other side of the solenoid, the wave function goes down exponentially. It's strange, but it is correct. 

\GH Do you remember what led you to discuss this particular case with him? 

\YA Dirac wrote on the magnetic monopole \citep{dirac1931monopole}. There he discussed phases around quantized flux. So, if he had thought carefully at that time he could have found the Aharonov-Bohm effect. I discussed that with him, and this is where I gave the example of the array of solenoids. That's what I remember about it. 

\GH What were you working on at the time? 

\YA Yeshiva is where I started to think seriously about measurement theory and came to the idea that we should change our understanding of what happens when you do a measurement. It occurred to me that if the measurement tells you not only information about the future, but also information about what was there before---maybe even before the measurement---the particle was already there when you found it. And then you have to reformulate quantum mechanics, saying that a particle at a given time has two kinds of information: one coming from the past measurement, and one coming back from the future.

That idea, which I shared with Bergmann and Lebowitz, led to what is called the ABL formula \citep{aharonov1964abl}. But that was the beginning of it. The idea was that the way to avoid the catastrophe that happens when you localize a particle—the only way to solve it—is by saying that the information is also relevant for the past, and not only for the future. Namely, I think that in the past the particle was described by two things: one that came from what had happened earlier, but also that already before you made the measurement, it somehow ``knew'' what was going to happen—what you were going to measure.

Altogether, it was a very exciting period, in terms of my own development and my collaborations. I was there altogether about eight years. Lenny Susskind was there most of the time. We did a lot of work together \citep{aharonov1967superselection,aharonov1967signchange}. Aage Petersen also came and joined me, and we developed the idea of modular momentum together with Hugh Pendleton.\footnote{See \cite{aharonov2025hetzroni}, \S6. }

\GH Around the same time, John \citet{bell1964} published his influential paper on correlations in the EPR settings. What did you think about it? 

\YA I did not find this very exciting. For me it was much more important to pay attention to the Aharonov-Bohm kind of nonlocality. In my view there are two types of nonlocality, EPR type, which is kinematical, and the AB type. With entanglement the probabilities of two-particle states are not the product of the probabilities of one-particle states. So somehow this difference is not as surprising as the difference in dynamics, which is much more important. So I thought that AB is more interesting because it has to do with a real, dynamical, change of the probability distribution. I still believe that entanglement is interesting, but it's not the one effect that demonstrates the most important new aspects of quantum mechanics. 

\GH At your time there you also met John Clauser, who later conducted the experiment to measure Bell correlations \citep{freedman1972bell}. 

\YA Yes, he was at that time at Columbia University,  and he came to meet me in Yeshiva. He said to me that he wanted to do an experiment to check entanglement in conditions where it will be nonlocal. And he asked me what I think would happen. So I told him, of course, it will be found out that quantum mechanics is right. He said, no. I believe that in this situation  quantum mechanics will not work. So we had a bet on one dollar, and he did perform the experiment a few years later. He  sent me the article, with the dollar  attached to it. 

\begin{figure}
    \centering
    \includegraphics[width=0.6\linewidth]{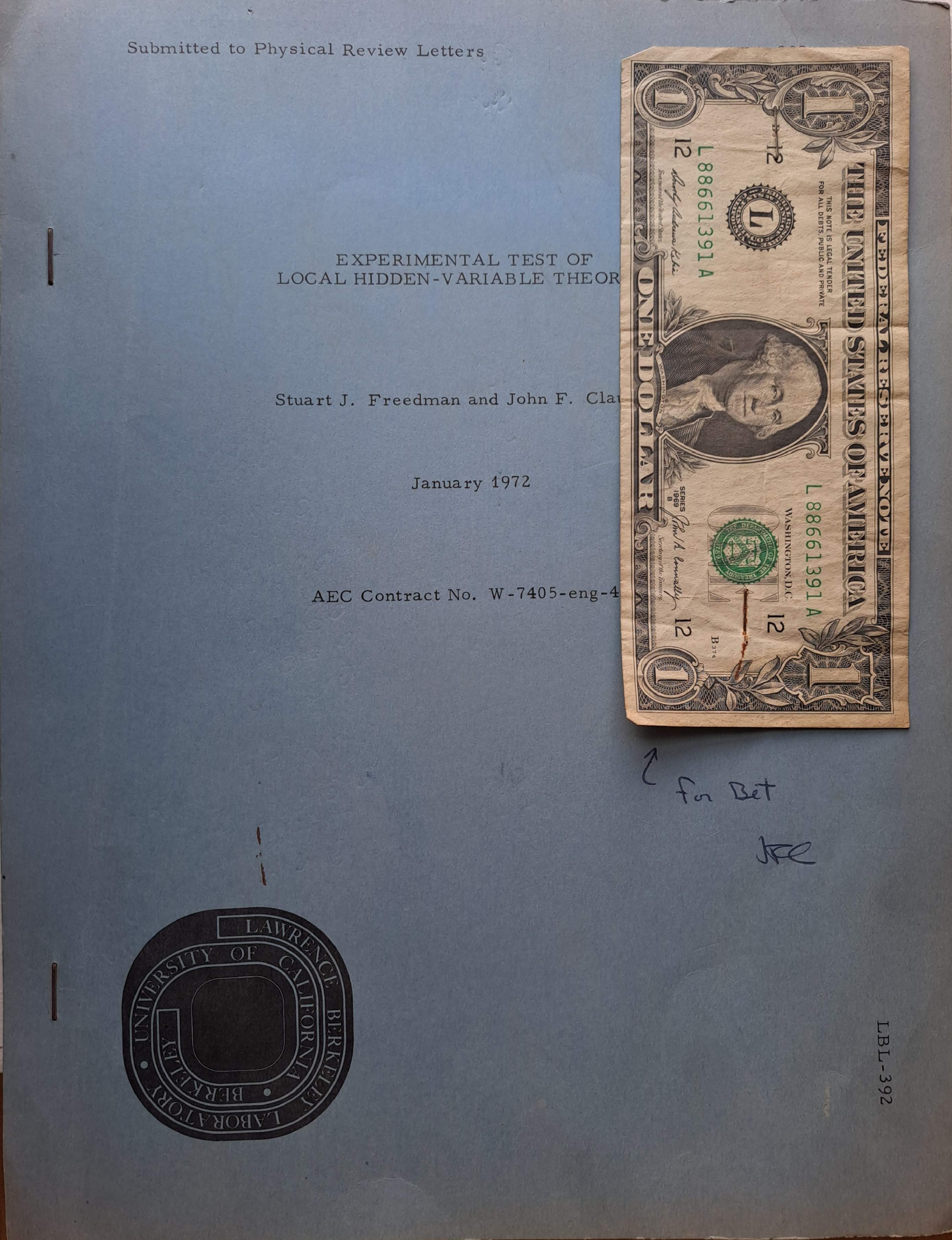}
    \caption{The copy of the submitted draft of \citet{freedman1972bell} that Clauser sent to Aharonov with the one dollar ``For Bet'' attached.  }
    \label{yang-aharonov-tonomura}
\end{figure}

\GH But why were you so sure that quantum mechanics would be right?

\YA  When you have particles close to each other, like in the case of an atom, you have an electron close to the proton, and obviously quantum mechanics is right. I couldn't see a situation where suddenly, at larger distances, quantum mechanics would be wrong. It didn't make sense.

\GH Was your confidence also related to your paper with Bohm on Wu's experiment, in which you did the calculations carefully against the actual experiment?

\YA Maybe. But then, of course, in the case of C. S. Wu's experiment, the condition of nonlocality, the spacelike separation, was not carefully checked. 

\GH Do you remember other discussions from the time of Bell's work, or your paper with Bohm on EPR? 

\YA Yes. With Abner Shimony. I met him many times and  we discussed this. He thought that \citet{aharonov1957epr}  was the most important article of mine with Bohm. He said that the understanding of this problem was my most important contribution. I told him I don't believe so. I told him why I believe that the nonlocality of the AB type is more important. He was extremely excited about entanglement, and I did not find it as exciting as he did.

\GH In an oral history interview \citep{shimony2002oralhistory}, Shimony  recalls a conversation where you told him that no experiment is needed to test Bell's correlation beyond your analysis of Wu's experiment. 

\YA Yes. Exactly for that reason, and because I felt very confident about the predictions of quantum mechanics. 

\GH In 1967 you started a position at Tel Aviv University. 

\YA Yes, after maybe seven years in Yeshiva University I met Yuval Ne'eman. That was right after the Six-Day War, and he convinced me  to join  Tel Aviv University. When I came to Tel Aviv, I think that other than Yuval, I was the only professor there. It was at the beginning of the department. I was in Tel Aviv until my retirement around 1993. I also soon started to have a second position at the University of South Carolina.

\GH How were you splitting the time after 1967 between Tel Aviv and the United States? 

\YA During these years I spent the summer in the United States, and maybe a month in the middle of the academic year, when there was a semester break. Occasionally I also spent my sabbatical in South Carolina. After I retired from Tel Aviv, I started to spend more time there, and later at George Mason University and then at Chapman University. 

\GH Let's go back to the late 1960s, when you joined the University of South Carolina. Who invited you there? 

\YA Ed Lerner, whom I first met when I was in Brandeis for the first time. I  first came to South Carolina just for one month's visit, it was in the spring. It was beautiful there. So I liked the place, and I decided to come there.

\GH Were there many people doing quantum physics in South Carolina?

\YA Ed Lerner was there. James Knight was also there, and John Safko. Soon people came to work with me. Dave Albert, Lev Vaidman. Jeeva Anandan later joined the department.

\GH How did you start working with David Albert?

\YA David Albert was doing his PhD in New York at the Rockefeller Institute with someone else. He found out about my work, came once to South Carolina and introduced himself. He was very interested in what I was doing, and very talented. I remember he started talking about information... So while he was doing his PhD in New York he also started to interact with me. After he graduated I got him a position in South Carolina, and he came there as an assistant professor. He was there with me two or three years. He also came to Israel with me.

\GH During these years you pioneered the conceptualization of quantum frames of reference \citep{aharonov1973carmi,aharonov1974carmi2}, and later in \citet{aharonov1984kauffer}. What was the origin of these idea and collaborations? 

\YA The idea, in fact, started earlier, I first discussed it with Aage Petersen during a year, or perhaps half a year, that I spent for a second time at Brandeis in the mid 1960s, when we were working on modular variables.  What I did with Petersen was really the beginning of quantum frames of reference. We did not publish it, I only gave lectures about it. The first publication on frames of reference was with Carmi. The beginning of the work I later published with my PhD student Tirzah Kaufherr was done with Petersen. 

\GH Were you in touch with Petersen when you were working on this in  the 1980s? 

\YA I think that when I wrote the article with Kaufherr, Petersen already had begun to develop Alzheimer’s disease. I learned about it from a mutual acquaintance, Emile Grgin, who previously had a temporary position at Yeshiva University and continued to work with Petersen. I think it was about ten years before Petersen died.\footnote{Petersen died in 1992; see \citet{halvorsonaagepetersen}. }

\GH You and Petersen do begin a paper on a different topic, spin and interference \citep{aharonov1967petersen}, by noting that an EPR state, in its original formulation using position and momentum, is in fact a constraint on the relative distance and total momentum. You also note  that they commute with each other. Was this also the starting point for quantum frames of reference? Can you describe how the idea evolved from your work with Petersen to your later papers?

\YA
I don’t remember how we began to think about it... It was later straightforward, more or less, to think about rotating frames, because when you go to a rotating frame, you see there something like a magnetic field and an electric field. So I knew about that, so it was quite straightforward to do the things we did with it.

\GH You knew Gideon Carmi already from the Technion, and you were together in Bristol and in Brandeis. 

\YA Yes, we talked a lot about physics all the time, but the only papers we wrote together were these two.

\GH Your papers make a connection between quantum frames of reference and the equivalence principle. This subject has recently gained lots of attention.\footnote{E.g. in \citet{bartlett2007referenceframes,giacomini2019qrfs,cepollaro2025entanglementcoherence} and references therein. } How did you make the connection? 

\YA It began earlier, in my paper with Daniel Wisnivesky \citep{aharonov1967wisnivesky}, where we were looking for a gravitational analogue of the AB effect.

\GH What made you later move to George Mason University in the 2000s, and later to Chapman? 

\YA Around 2000 I spent some time in Boston University. Jeff Tollaksen was my PhD student there. A few years later he was at George Mason University in Virginia, and arranged a meeting for me with Daniele Struppa, I think he was a dean there. Daniele invited me first to give a series of lectures about my work. That worked very well, so he invited me to stay there. Later Daniele became the provost at Chapman University, and then he brought me along there. 

\GH I also wanted to ask you about your teachings. Among other topics, you gave in Tel Aviv University a course on quantum foundations. You later wrote a non-standard textbook about it with Daniel Rohrlich \citep{aharonovrohrlich2008QP}. 

\YA The book relates to my ideas, but in my actual teachings I never prepared my lectures in advance. It was always based on just the spur-of-the-moment ideas. My teaching was quite different from the usual courses. My aim was not to try and teach facts, but to try and teach how to think. I think it went very well. I always got good feedback.  

\GH What are the important ways of thinking that you wanted your students to acquire? 

\YA Even when I taught, say, classical electromagnetism or basic quantum mechanics, I was always trying to find some way to introduce a new subject as a paradox or a question, not just presenting facts, but trying to see what could be done about some problem. I tried to get the students to see by themselves how to develop the subject.

\GH During your decades in Tel Aviv and also in Yeshiva University, University of South Carolina, George Mason University, and Chapman University, were you involved in any form of management or did you take part in committees in these institutions? 

\YA
No. I  never did any kind of administrative work. I convinced people that I can’t do it. I did my research, I taught, and I had PhD students. 

\GH
Yet, you probably were involved in recruiting people or recommending people to recruit?

\YA
People came to Tel Aviv University and to Chapman because of me, but I was never involved directly in trying to convince them to come.

\section{The road to new forms of measurements}

\noindent\fbox{%
\begin{minipage}{\dimexpr\textwidth-2\fboxsep-2\fboxrule\relax}
\footnotesize
\textbf{Box 1. Weak measurements, weak values, the quantum Cheshire Cat, and protective measurements}
A standard von Neumann measurement correlates an observable $A$ with a pointer coordinate $Q$ using an interaction Hamiltonian:
\begin{equation}
\label{WeakMeasurementCoupling}
H_{\rm int}(t)=g(t) A P,\qquad [Q,P]=i\hbar,\qquad \int g(t)\,dt=g_0 .
\end{equation}
The function $g(t)$ is the measurement pulse: it is non-zero only around $t=0$ while the apparatus interacts with the system. If the system is in an eigenstate $A|a\rangle=a|a\rangle$, the pointer shifts by $g_0a$. A \textbf{\emph{weak measurement}} uses an initial pointer wavefunction whose width is large compared with the shifts $g_0 a_i$ associated with the eigenvalues of $A$; the shifted pointer states therefore strongly overlap, so a single run gives little information and causes only small disturbance. Averaging over many equally prepared runs recovers the shift.

A \textbf{\emph{weak value}} is associated with a weak measurement (performed with a pointer whose initial width is sufficiently large) between two projective (standard) measurements that define an ensemble in terms of a pre-selected state $|\psi\rangle$ and a post-selected state $|\phi\rangle$. The shift in this case is the  weak value
\begin{equation}
\label{WeakValue}
A_{\rm w}=\frac{\langle \phi|A|\psi\rangle}{\langle \phi|\psi\rangle},
\qquad
\Delta Q=g_0\,{\rm Re}\,A_{\rm w}.
\end{equation}
Because the denominator may be small, $A_{\rm w}$ need not lie between the eigenvalues of $A$. For spin, pre-selecting a spin direction $|\xi\rangle$ at an angle $\alpha$ with $+x$ (on the $x$--$z$ plain) weakly measuring $\sigma_z$, and post-selecting $|+x\rangle$ gives the value:
\begin{equation}
\label{SpinWeakValue}
(\sigma_z)_{\rm w}
=
\frac{\langle +x|\sigma_z|\xi\rangle}{\langle +x|\xi\rangle}
=
\tan\frac{\alpha}{2}.
\end{equation}
Thus, a value of $\alpha$ close to $\pi$ can give $(\sigma_z)_{\rm w}=100$. A small coupling, such as a weak Stern-Gerlach deflection, is then amplified by this factor, at the price of rarer successful post-selection.

\vspace{0.4em}
\noindent
\begin{minipage}[t]{0.65\linewidth}
\vspace{0pt}
\textbf{\emph{The quantum Cheshire Cat experiment}} applies the same rule to a photon in a two-arm interferometer. Prepare and post-select the states:
\[
|\Psi\rangle=\frac{1}{\sqrt2}(i|L\rangle+|R\rangle)|H\rangle,
\qquad
|\Phi\rangle=\frac{1}{\sqrt2}\bigl(|L\rangle|H\rangle+|R\rangle|V\rangle\bigr).
\]
Let
\[
\Pi_L=|L\rangle\langle L|,\qquad
\Pi_R=|R\rangle\langle R|
\]
be the path projectors, and let
\[
\sigma_z=|+\rangle\langle +|-|-\rangle\langle -|
\]
be the circular-polarization operator ($|\pm\rangle\equiv(|H\rangle\pm i|V\rangle)/\sqrt2$). The polarization in each arm is then represented by
\[
\sigma_z^{(L)}=\Pi_L\sigma_z,\qquad
\sigma_z^{(R)}=\Pi_R\sigma_z,
\]
where the tensor product with the identity on the other degree of freedom is implicit. The weak values are
\begin{equation}
\label{CheshireWeakValues}
(\Pi_L)_{\rm w}=1,\quad
(\Pi_R)_{\rm w}=0,\quad
(\sigma_z^{(L)})_{\rm w}=0,\quad
(\sigma_z^{(R)})_{\rm w}=1 .
\end{equation}
The resulting picture is therefore: the photon, or ``cat'', is in the left arm, while its circular polarization, the ``grin'', is in the right arm. Since the measurements are weak, these two weak probes can be applied in the same experiment.
\end{minipage}\hfill%
\begin{minipage}[t]{0.35\linewidth}
\vspace{0pt}
\centering
\includegraphics[width=\linewidth]{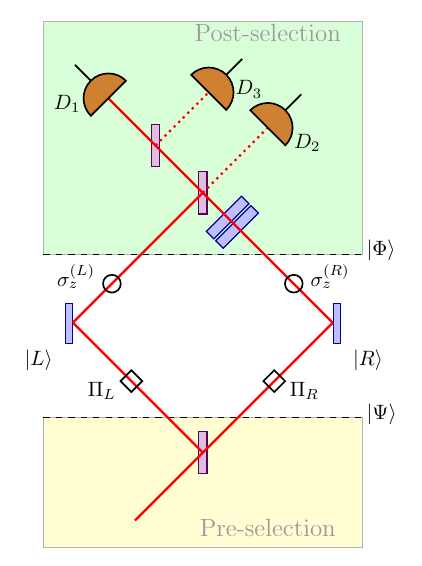}
\end{minipage}

\vspace{0.4em}\textbf{\emph{Protective measurements}} ask a different question: can a single quantum system reveal an expectation value? Here the system is protected in a non-degenerate energy eigenstate $|\psi_n\rangle$ and the coupling in Eq.~\eqref{WeakMeasurementCoupling} is applied weakly and slowly ($g(t)\simeq g_0/T$ over a sufficiently long time
$T$), 
namely, the state $|\psi_n\rangle$ is an eigenstate of the free Hamiltonian  and the interaction acts as a small perturbation over a long period rather than
impulsively. In this case
the adiabatic theorem keeps the state unchanged while
\begin{equation}
\label{ProtectivePointerShift}
Q(T)-Q(0)=g_0\langle\psi_n|A|\psi_n\rangle .
\end{equation}
Although the instantaneous coupling is weak, the accumulated shift can exceed the
pointer width. Thus the pointer reads an expectation value from one protected system, not an ensemble average. Choosing $A=\rho_D(x-x_0)$, with $\rho_D\to\delta(x-x_0)$, gives $\langle A\rangle\to|\psi_n(x_0)|^2$; suitable current measurements give the phase gradient. In this sense, protective measurements can probe the wavefunction itself, provided the protecting gap or monitoring is part of the experimental arrangement.

\end{minipage}}

\GH
You collaborated over the years with many people, some of them were really long-term collaborations, and some of them are now well known. We mentioned David Albert and Lev Vaidman. Later there were also Jeeva Anandan, Daniel Rohrlich, Sandu Popescu, Avshalom Elitzur, and others. How did your collaborations work? 

\YA
I actually always liked to work alone. But when I was a professor at Tel Aviv, I couldn’t refuse to accept students for a PhD if they came to me. But I always tried my best to convince them that they should not work in this field, because at that time quantum foundations was not a popular field. It came into fashion later, when people started to worry about quantum gravity and about quantum computers. But at that time, it was very unlikely for people to find a position if they worked on foundations.

So each time people came to me and asked to do their PhD, I tried to talk them out of it, to make sure that they understood how difficult it would be later to get a position. So in general, people who worked or did their PhDs with me were committed to foundations. They said that if not for foundations, they would not do physics. That was the only condition for me to accept them. 

I don’t remember the exact years... In the States, there was Daniel Wisnivesky in Yeshiva. Later Susan D’Amato , Chi-Kwan Au, and Hsiang-Wu Huang. They did their PhDs with me in South Carolina. And then Dave Albert. Jeeva Anandan was  a physics professor at the University of South Carolina, we collaborated there from the 1980s until his death of Lou Gehrig's disease, more than 20 years ago. 

In Tel Aviv I think Lev Vaidman was first. Sandu Popescu later.  Daniel Rohrlich joined even later, he already had a PhD when he came to Israel.  
I met Avshalom Elitzur at a meeting to which I was invited where Yeshayahu Leibowitz was present. There was a young fellow there asking interesting questions, and afterward I found out that he came from a religious family, had no formal education, but had learned physics by himself through extensive reading. I invited him to my office, and after a number of meetings I decided to accept him as my PhD student even though he did not have a university degree; I got a special permission from the university rector or president.

\GH A connecting thread among many of your papers, from your PhD to recent years, is the attempt to understand quantum theory by asking: `what could, in principle, be measured?'. Two  concepts that became particularly influential, weak measurements and weak values (see Box 1),  were introduced in \citet{aharonov1988weak}. One idea on which this paper was based is the idea of what happens to the particle between two measurements, and the time-symmetric ABL formalism \citep{aharonov1964abl} on which you elaborated in \S\ref{sixties-seventies}. Can I ask why, with this idea in your hands since 1964, didn't the idea of weak measurements come up earlier than the late 1980s? 

\YA I admit that at that time in the 1960s, when I found that modular momentum plays a role, I felt that I already had a good explanation of interference, in terms of this operator that describes a single particle rather than ensemble. Then, I felt unhappy about the idea of, say, knowing through which slit the particle goes, because that seemed completely uncertain. I did not think about the possibility of doing a which-path measurement for an ensemble of many particles without disturbing each one, only later finding out which way it went. That idea came much, much later.

\GH The other idea on which weak values are based is weak measurements. A weak measurement is basically a von Neumann measurement in which the pointer ends up in non-orthogonal states, namely the pointer displacement is much smaller than its initial uncertainty. How did this idea develop? 

\YA I looked at the simplest example, a spin-half particle, no magnetic field, so the components are constants of motion. And at time $t_1$, you measure and find $\sigma_x=1$. Then, at a later time $t_2$, you find $\sigma_y=1$. In the ABL approach during the time between $t_1$ and $t_2$, we knew not only $\sigma_x$, also $\sigma_y$. Then I asked what kind of measurement can I do in the time between these two measurements that will show that both $\sigma_x$ and $\sigma_y$ were equal to one? At first I thought, `why don't I measure $\sigma_x+\sigma_y$'? That's a single measurement, and that should give the outcome $2$, if each one of them is $1$.

But very quickly I realized that this sum is just  $\sqrt2$ times the component of spin in the 45-degree direction, which can be either one or minus one, but never square root of two. So at first I got alarmed and thought `my God, the picture doesn't seem to work'.

But then I thought more carefully, and I realized that when I try to measure $\sigma_x + \sigma_y$, that measurement necessarily makes $\sigma_x$ itself and $\sigma_y$ itself uncertain. To make the measurement, I have to put an inhomogeneous magnetic field in the 45 direction. That means that the particle experiences an uncertain rotation around that direction. So that measurement necessarily changes both $\sigma_x$ and $\sigma_y$. So it's not surprising that I could not see the sum to be two.

That's when I started to think: what will happen if I make a measurement that interacts so weakly with the particle that it will not change $\sigma_x$ and $\sigma_y$? The obvious worry was that such a measurement would give me no information. But then I realized that if the information is proportional to $\epsilon$, the disturbance to the system is proportional to $\epsilon^2$. Therefore, in an ensemble of  $1/\epsilon$ particles  I will be able to see the effect of the measurement, and I will indeed find the value two. So that was the beginning of thinking about weak measurements.

The next step was when I found the magic formula, it’s a simple and extremely useful formula that does not usually appear in textbooks.  If I have an ensemble of particles, all of them are described by the same state, $\psi(x)$, and I can make a measurement of any Hermitian operator $A_n$ that acts on the $n$-th particle. The idea is to take an operator $A$ acting on the ensemble, which is the sum of $A_n/n$. The result of the operator $A$ acting on $\psi$ has a component proportional to $\psi$ and a component orthogonal to it. The magic formula says that the proportionality constants are the expectation value and the uncertainty in $A$, namely  $A\psi=\langle A\rangle\psi+\Delta A\psi_\perp$. I used this formula to prove that in the limit that n goes to infinity, the ensemble is an eigenvector of $A$  with an eigenvalue that is the expectation value of $A$. There is a correction that goes to zero when $n$ goes to infinity. So if I take, for example, the sum of projection operators for position x, that gives the expectation value, $\psi^*(x) \psi(x)$. And I can therefore find the state of each particle, if they are all in the same state, without disturbing any of them, just as long as I have enough of them. 

\GH Why did you choose the title `How the Result of a Measurement of a Component of the Spin of a
Spin-1/2 Particle Can Turn Out to be 100'? It doesn't seem to reveal the actual internal logic of weak values. 

\YA At that time I thought that this provocative title would make people listen, but the first reaction to this article was only a number of years later. At the beginning nobody paid attention. What it shows is that the weak value, because the denominator can be very small, is a way to amplify things. For example, you take many photons with some definite polarization. And you send them through matter that changes their polarization just by a tiny  amount that you can never directly measure. But then if you take the whole ensemble and bring it through another polarizer that is nearly orthogonal to the original one, most of the photons will not go through. With the small fraction that goes through, you can see the weak value. For example, John Howell, when he was still at Rochester, did such an experiment \citep{dixon2009beamdeflection}, he said that he was able to see a small shift in the direction, such that if he sent the beam to the moon, at the moon it would have shifted by only the thickness of one hair. This kind of amplification of very small effects is now used in many laboratories.

\GH The idea of time symmetry that started in the ABL paper \citep{aharonov1964abl} continued to develop, mainly in your papers with Lev Vaidman \citep{aharonov1990properties,aharonov1991completedescription}, to become the two-state vector formalism, and later also an interpretation \citep{aharonov2002tsvf,aharonov2007tsvfUpdated,aharonov2014measurementtsvf}. What were the major milestones in this development? 

\YA Once I discovered weak measurements, it led to a completely new reality because those measurements don't change the system, and therefore they tell you what is really there. So once I had this formula for the weak values, I saw immediately that it leads to completely new and interesting aspects of pre- and post-selected quantum systems.  

One important application was the case of negative kinetic energy. This is for a particle bound in a potential well, i.e. with negative total energy, that is post-selected in a position outside the well, where the potential energy is zero. The weak value of the kinetic energy is exactly the negative total energy \citep{aharonov1993negative}. 

Another example from the same time was the quantum random walk \citep{aharonov1993quantumrandomwalks}. Suppose a particle with spin one-half, you arrange an experiment where the particle moves a small distance $l$ to the right if it's $\sigma_z$ is equal to $+1$, and to the left if it is $-1$. So if you start with a state which the spin is in the $x$ direction, you know that you have exactly a probability one-half that it will move to the right and a probability one-half that it moves to the left. You arrange that each step is much smaller than the width of the wave function. It turns out that the displacement after $N$ steps can be much larger than the classical value $Nl$. 

Then there were super-oscillations.\footnote{\citet{berry1992superoscillations}, and see also \citet{aharonov2025hetzroni}, \S7. } Later the quantum Cheshire Cat \citep[see also Box. 1]{aharonov2013quantumcheshirecats} and the quantum pigeon-hole principle. These examples, and some others contributed to the understanding of post-selection and weak values and their place in the interpretation of the theory. Basically, the way the post-selection divides the pre-selected ensemble into sub-ensembles is the basis of the two-state-vector formalism. 

\GH As a side note, some people see in the name of the `two-state vector formalism' a political pun, referring to the `two-state solution' of the Israeli-Palestinian conflict. Is this intended? 

\YA No, not at all. I never thought of that. 

\GH Long before the quantum random walk paper and the quantum pigeonhole paper, in 1993 with Lev Vaidman,  you came up with the idea of protective measurements \citep{aharonov1993protective1,aharonov1993protective2}. These are measurements that probe the quantum state of a single system while protecting the state from changing using an appropriate Hamiltonian. How did this idea start? 

\YA I don't remember the origin of this idea,  but it is another version of weak measurements. With protective measurements you interact very weakly and also with a very slow change. You introduce a weak coupling of a system to a Hermitian operator $A$, and if the duration of the measurement goes to infinity and the strength goes to zero, you end up just measuring the average of $A$ without changing the state, which is protected. 

Weak measurements relate to an ensemble of particles. We can think of a protective measurement as  an ensemble of measurements over the same particle, during the very long time of the measurement the particle returns many times to its original state. If you think about the protecting potential as a series of repeated measurements, it is the quantum Zeno effect. In this sense protective measurements are another version of weak measurements, effectively like an ensemble. 

\GH From the papers it seems that you have changed your mind about this, at the beginning there was a lot of emphasis on measurement being performed over a single particle. 

\YA Yes. At the beginning I thought that finally I can see that it's possible to associate a wave function with a single particle. But later I realized that's wrong because, really, it is effectively an ensemble. I don't remember what made me realize this. You can associate $\psi$ with a single particle, but then you can never find out which $\psi$ it is, because if you have two $\psi$'s that are not orthogonal, you cannot distinguish between them with a single particle. 

\GH You also mentioned the quantum Cheshire Cat (see Box 1). What do we learn from that experiment? 

\YA I chose this name, the quantum Cheshire Cat, because I read Alice in Wonderland when I was in high school, and thought that the idea of the smile remaining after the cat disappeared was peculiar, it seemed  a natural name for this idea of separating the particle from its spin. What led me to this idea? Well, I think I just played with different pre- and post- selections. And then I discovered that I can actually do so by starting with $\sigma_x$ here superposed with $\sigma_x$ there, and ending up with $\sigma_x$ here and the opposite $\sigma_x$ there, playing with that I saw that I can actually separate the spin from the particle. It is an experiment that people soon performed in a lab \citep{denkmayr2014quantumcheshirecat}.  
\GH And then, the quantum pigeon-hole paper \citep{aharonov2016quantumpigeonhole} is another example of the peculiarities of pre- and post-selection. It shows how this simple combinatorial principle, stating that if there are three pigeons in two pigeonholes then at least two of the pigeons must be in the same hole, is violated for ensembles defined by a quantum pre- and post-selection. The paper won the National Academy of Sciences' Cozzarelli Prize for a paper published in 2016. 

\YA The original idea was actually about spins: you start with $\sigma_x=1$ and end with $\sigma_y=1$. In this case the weak value of $\sigma_z$ is $i$, it's imaginary. Now you have two particles, separated from each other, but identical in the same pre- and post-selected spin states. They are not entangled, so the weak value of the product equals the product of the weak values, so the weak value of the product of the $z$ component of the spin equals $-1$. You can have as many particles as you like, and the product of the $\sigma_z$ of any two of them would be $-1$. Now, consider two boxes where the particles could be, and replace  $\sigma_z$ with an operator measuring in which of the two boxes the particle is. These operators, in the case of three particles, give you the paradox. For two particles it's not a problem that the product is $-1$, the two particles are not in the same box. But with three particles, every two particles are not in the same box, even though there are only two boxes.  

\GH Among the authors, in addition to Sandu Popescu, Daniele Struppa, and Jeff Tollaksen, whom you mentioned before, are Fabrizio Colombo and Irene Sabadini. 

\YA Yes, they are Italian mathematicians, acquaintances of Daniele. They also wrote on superoscillations \citep{aharonov2017superoscillations}, and other things. They did a lot. 

\GH I have two questions to conclude with. First, what do you consider the most important point you want people to know about your work? 

\YA Beyond  pushing the idea of the ABL and the two state vectors, I want also to push the idea that the Heisenberg approach to quantum mechanics is better than the \schrodinger\ approach. 

\GH Are these two views related to each other? 

\YA Consider a particle that is initially in a superposition of being in the left box and being in the right box. Then you open the right box and you find it there. What happens in between? In the \schrodinger\ picture we know two things about the particle, from its destiny and from its history. We know with certainty that it's in the right box,  and  that with certainty it is half in the right box and half in the left box. This is a very bad language, because these are two contradictory statements. The Heisenberg picture removes the contradiction. It tells you it was always in the right box, and that it also has another property that is a nonlocal property that can allow it to move between the boxes. This is a much better language. 

\GH And the second question: you have been working on quantum mechanics since the 50s, that's about 70 years now. What has changed in the community's understanding of the field? 

\YA Do you remember the story that I told about Nathan Rosen?\footnote{See \citet{aharonov2025hetzroni}, beginning of \S2. } He told me that to work on foundations is for old people. Young people should not worry about those things, because it doesn't lead to anything... Nowadays, people know that work in foundations is very important, for quantum computers and also for understanding how to combine quantum mechanics and general relativity. So certainly, it's much more respected today to work on foundations than it was when I started.

\section*{Acknowledgements}
GH would like to thank Michelle Frank, Stefano Furlan, Judy  Kupferman, Jean-Philippe Martinez, Marij van Strien,  Daniel Wisnivesky. This research was supported by the Open University of Israel Research Fund (grant no. 515997).

\end{document}